\documentclass[12pt,preprint]{aastex}

\slugcomment{ }

\shorttitle{Hypercompact HII regions}
\shortauthors{Eric Keto}

\begin{document}

\title{The Early Evolution of Massive Stars: Radio Recombination Line Spectra}

\author{Eric Keto }
\affil{Harvard-Smithsonian Center for Astrophysics, 60 Garden St., Cambridge, MA 02138}
\and
\author{Qizhou Zhang }
\affil{Harvard-Smithsonian Center for Astrophysics, 60 Garden St., Cambridge, MA 02138}
\and
\author{Stanley Kurtz }
\affil{CRyA, Universidad Nacional Aut\'onoma de M\'exico, Apdo. Postal 3-72, 
       58090 Morelia, Michoac\'an, Mexico}

\begin{abstract}
Velocity shifts and differential broadening 
of radio recombination lines
are used to estimate the densities and velocities of the ionized gas
in several hypercompact and ultracompact 
HII regions. 
These small HII regions
are thought to be at their earliest evolutionary phase and associated with
the youngest massive stars.
The observations suggest that these HII regions are
characterized by high densities, supersonic flows and steep density
gradients, consistent with accretion and outflows 
that would be associated with the formation of massive stars.

\end{abstract}


\section{Introduction}

Massive star forming regions are marked by small ($\leq 0.1$
pc) HII regions \citep{Gaume1995b, Depree1997, Tieftrunk1997, Depree1998, 
Wilson2003, Depree2000, Depree2004, Depree2005} 
that are characteristically different from many of the
few thousand ultracompact HII (UCHII) regions scattered throughout the
galaxy \citep{WoodChurchwell1989, Kurtz1994, Walsh1998, Giveon2005a, Giveon2005b}.  
First, these very small HII regions have electron
densities that are higher ($n_e \geq 10^5$ cm$^{-3}$) than those
of UCHII regions ($n_e\sim 10^4$ cm$^{-3}$)
\citep{Kurtz2000, KurtzFranco2002, Pratap1992}. 
It is unclear how high the densities might be, because 
the density cannot be reliably determined from the radio continuum
if the gas is optically thick.
Second, when
observed at centimeter wavelengths, these small HII regions often show
very broad radio recombination line (RRL) widths, more than 3 or 4 times the
thermal line width \citep{Altenhoff1981, Zijlstra1990, Afflerbach1994,
Depree1994, Gaume1995a, Depree1995, Depree1996, Depree1997,
Keto1995, Johnson1998, JaffeMP1999, Keto2002a,
Sewilo2004, KetoWood2006}.  Not all the small HII regions have this
property, but those that do are sometimes referred to as broad
recombination line objects.  The breadth of the RRLs could be caused by a 
combination of electron impact  (pressure) broadening associated 
with high gas density plus broadening caused by spatially unresolved 
gas motions. However,
the relative contributions
of broadening due to gas pressure and gas motions are not known.
Third, these HII regions often have
continuum spectral energy distributions (SEDs)
that are roughly linear with frequency
over decades in wavelength from the centimeter to the submillimeter
\citep{Hofner1996, Franco2000, Testi2000, Beuther2004, Ignace2004}.
The underlying cause of this scaling is not known although spatially
unresolved density
variations are a possibility.

These small and dense HII regions are sometimes referred to as 
hypercompact HII (HCHII) regions.  While the term
hypercompact implies a size smaller than ultracompact (UC), the three
properties of high electron densities, broad recombination lines, 
and linear SEDs, are also seen in some UCHII regions
($\sim 0.1$~pc).  The common characteristic between these unusual
UCHII regions and the HCHII regions might be that
they are all young and
at the earliest evolutionary stage, but the 
unusual UCHII regions might
contain more or brighter stars than their smaller HCHII counterparts.
The HCHII and unusual UCHII regions represent 
some of our best sources of information on the processes in the formation of 
massive stars, and this motivates our understanding of their special properties.

In this paper we report on multi-frequency RRL observations that
address the origin of the three properties of the small HII regions
and the processes of massive star formation.  Because 
broadening due to pressure is dependent on frequency whereas broadening due
to gas motions is independent of frequency, the widths of RRLs
observed at more than one frequency can be used to separate the
contributions of each broadening mechanism. Since the pressure
broadening is proportional to the density, this separation allows a
determination of the electron density. Similarly, 
since the broadening due to gas motions is proportional to
the range of velocities, 
we can also measure the magnitude of the gas velocities.

\section{Observations}\label{observations}

We selected a few well-studied, very bright, very small HII regions 
that previous observations suggest have one or more of the 
three properties of the small HII regions. For each source
we have one observation of the H30$\alpha$ line made with the 
Submillimeter Array (SMA) and one or more observations of
centimeter wavelength RRLs made with
the National Radio Astronomy Observatory
Very Large Array (VLA)\footnote{The National Radio Astronomy
Observatory is a facility of the National Science Foundation
operated under cooperative agreement by Associated Universities,
Inc.} 

The SMA observations of the H30$\alpha$ line (231.900959 GHz) in the
HII regions G10.6-0.04, NGC7538-IRS1, W51e2, and G28.20-0.04 were made
in 2005 September with a spectral resolution of 0.83 km~s$^{-1}$, a
bandwidth of 2 GHz, and an angular resolution of 1.0$^{\prime\prime}$.
Observations of G45.07+0.14 were made in 2005 October with an angular
resolution of 0.4$^{\prime\prime}$.  The noise level obtained in each
observation was 60 mJy beam$^{-1}$ channel$^{-1}$.  Analysis of the data
indicated that some of the baselines were not well-defined, hence the
positions and flux densities of the sources are uncertain.  
The standard deviation of the flux of 3C454.4, measured once for each
observation,  is
39\% of the average measured value
(table \ref{millimetercalibration}).
If we exclude the low measurement of 3C454.4 corresponding to 
the G45.07+0.14 data,
then the standard deviation falls to 14\%
of the average.  Our positions from the SMA agree with those of
the VLA to better than the angular resolution.  The HII regions 
are unresolved by the SMA observations,
and hence the widths of their spectral lines are unaffected by baseline
errors.  The SMA data were processed in the SMA data reduction package
MIR, and in MIRIAD.

Most of the VLA observations were made in 2003 and 2005. 
We reprocessed observations of the
H76$\alpha$ line in G45.07+0.14 \citep{Garay1986} from the VLA
archives.  The observations of the H66$\alpha$ line in G10.6-0.4 were
previously reported in \citet{Keto2002a}. The other observations have
not previously been published.  Details of the observations are
presented in table \ref{VLAobs}. The VLA data were processed in AIPS;
calibration data are given in table \ref{centimetercalibration}.
The recombination line spectra are shown in figures 1 to 5, and
the  line widths and velocities based are given in table
\ref{CMrrl}.  

\section{Gas velocities, pressure broadening, and electron densities}\label{derivation}

Radio recombination lines in HII regions are broadened by three contributions:
thermal broadening, dynamical broadening due
to spatially unresolved motions in the HII region including both ordered flows
and turbulence, and pressure broadening
due to high electron densities.
We seek to estimate the separate contributions of each 
of the three broadening components by comparing the widths of RRL 
observed at different frequencies. This is possible because the 
width due to 
thermal and dynamical broadening, is independent of frequency
while the width due to pressure broadening 
decreases with frequency as $\Delta\nu\sim\nu^{-4}$ 
\citep{BrocklehurstSeaton1972, Griem1974}. 
Since the dynamical broadening is proportional to the gas velocity while
the pressure broadening is proportional to gas density, from a pair of RRL observations
we can estimate the velocities and densities in the ionized gas.

We determine the contributions of each of the three broadening
mechanisms as follows. We assume that the dynamical broadening, $\Delta\nu_D$, 
and the thermal broadening, $\Delta\nu_T$,
combine in quadrature,
\begin{equation}
\Delta\nu_G = \sqrt{\Delta\nu_D^2 + \Delta\nu_T^2}. 
\label{eq:equation1}
\end{equation}
This width due to gas
motions, both thermal and spatially unresolved ordered flows and turbulence,
combines with the Lorentzian width, $\Delta\nu_L$, due to
pressure broadening 
to produce a Voigt profile with width, $\Delta\nu_V$.
We approximate the width of the Voigt profile
by an algebraic expression 
\citep[][ equation 2.72]{Olivero1977, GordonS2002},
\begin{equation}
\Delta\nu_V = 0.5343\Delta\nu_L + [\Delta\nu^2_G + (0.4657\Delta\nu_L)^2]^{1/2}
\label{eq:voigtApprox}
\end{equation}
where $\Delta\nu_V$ is the full-width at half-maximum (FWHM) of the
Voigt profile, $\Delta\nu_L$ the FWHM of the Lorentzian, and
$\Delta\nu_G$ the FWHM of the Gaussian.  Because of the rapid decrease
in pressure broadening with frequency, the width of an RRL at a high
enough frequency must be  due solely to thermal broadening 
and spatially
unresolved gas motions.
Therefore we assume that this broadening,
$\Delta\nu_G$ in equation \ref{eq:voigtApprox}, is given by the
observed width of our highest frequency RRL, H30$\alpha$. 
If we compare the width of the H30$\alpha$ line to that of a lower
frequency line,
any increased width observed in the lower frequency line
must be due to pressure broadening 
because the width $\Delta\nu_G$ 
is independent of frequency 
We can then calculate the width due to pressure broadening,  $\Delta\nu_L$,
by solving the 
quadratic equation \ref{eq:voigtApprox}.

The range of gas velocities is estimated from equation (\ref{eq:equation1}).
Assuming an ionized gas temperature of 8000 K,
the dynamical widths are
15.3, 8.3, 27.1, 18.7 and 53.9 kms$^{-1}$
for G10.6-0.4, G28.20-0.04N, G45.07+0.14, W51e2, and
NGC 7538-IRS1 respectively.

The electron density may be estimated from the ratio of $\Delta\nu_L$ and $\Delta\nu_T$.
For $\alpha$ ($\Delta$N
= 1) transitions, the ratio of the Lorentzian width, $\Delta\nu_L$, resulting
from pressure broadening, and the thermal width, $\Delta\nu_T$, is given
by \citep{Keto1995},
\begin{equation}
{{\Delta\nu_L}\over{\Delta\nu_T}} = 
1.2 \bigg( {{n_e}\over{10^5}} \bigg)\bigg({{N}\over{92}}  \bigg)^7
\label{eq:ratio}
\end{equation}
where  $n_e$ is the electron
density in cm$^{-3}$, and $N$ is the principal quantum number.

As an example, we calculate the electron density in
G10.6-0.4 from the H30$\alpha$ and H92$\alpha$ RRL observations as
follows.   We use
the width of the H30$\alpha$ line, 24.5 kms$^{-1}$ (table \ref{CMrrl}) to estimate
$\Delta\nu_G$ at the frequency of the H92$\alpha$ line.  
Thus $\Delta\nu_G = 0.678$ MHz.
The measured width of the H92$\alpha$ line
is $\Delta\nu_V({\rm  H}92\alpha) = 1.03$ MHz. 
With these values and equation \ref{eq:voigtApprox}, the estimate of the
width due to pressure broadening  is $\Delta\nu_L = 0.578$ MHz
or 20.9 kms$^{-1}$. 
The thermal width, assuming a temperature of 8000 K, is $\Delta\nu_T =
1.43$ MHz or 19.1 kms$^{-1}$. The width due to 
gas motions
is $\Delta v_D = \sqrt{24.5^2 - 19.1^2} = 15.3$ kms$^{-1}$. 
The electron density from equation \ref{eq:ratio} is 
$9.1\times 10^4$ cm$^{-3}$. Similarly, from the observations of
the other centimeter RRLs observed in G10.6-0.4,  the electron densities are
$7.8\times 10^5$ cm$^{-3}$ from the H66$\alpha$ line and $2.5\times
10^6$ cm$^{-3}$ from the H53$\alpha$ line. Comparing these three
RRL, the electron density increases with the RRL
frequency.  If there is a density gradient in G10.6-0.4 this would
suggest that the higher frequency lines are seeing deeper into the
nebula where the electron densities are higher.

\subsection{Uncertainties}

We can check our approximation of negligible pressure broadening
in the H30$\alpha$ line from \ref{eq:ratio}.  
For the H30$\alpha$ line, this ratio is less than one as long as
the density is less than $3\times 10^8$ cm$^{-3}$. Thus our
assumption of negligible pressure broadening in the H30$\alpha$
line is valid up to this density.

The lines in our sample are not
consistently  Gaussians or by Lorentzian. One possibility
is that the asymmetric line profiles indicate asymmetries in
the gas motions in the HII regions,  for example, ordered flows rather than
isotropic turbulence. 
We determine the line widths by fitting Gaussians because
this provides a reasonable estimate if the
observed profiles
differ from a Gaussian most strongly in the wings and 
are similar in the core where the width is measured. 
A comparison of the line widths derived by fitting Gaussians
to Voigt profiles provides  an informal
estimate of the possible error introduced by this method. If
the Voigt profile is composed of a Gaussian and a Lorentzian
of equal FWHM, then a Gaussian fit overestimates the 
true FWHM of the Voigt by 10\%. 

We find that the procedure of using Gaussian line widths in equation
\ref{eq:voigtApprox} to estimate the pressure broadening 
is a more reliable method than some alternatives. For
example, one could imagine fitting a Voigt profile directly to 
an observed low frequency line to determine simultaneously both
the Gaussian and Lorentzian components. However, we find
this unreliable because the fitting procedure assigns the line width
to either the Gaussian or Lorentzian components primarily on the
shape of the line wings where the profiles differ most strongly, but
the signal-to-noise ratio is the weakest. Alternatively, one might
estimate the FWHM directly from the 3 channels with
the peak and half-power emission. However, because this estimate 
relies on  single channel measurements it is less
reliable than a line-fitting procedure that uses measurements at
all the channels across the line profile.  

Several lines of evidence, discussed later in the
paper suggest that there are density variations within the HII
regions.  Because all the HII regions in our sample are unresolved,
the measured electron density represents the average within the
HII region. 
The size of the HII region within the beam
(filling factor) does not affect this estimate of the 
average density
because the electron densities are
determined from the line widths rather than the emission
intensity. However, the unknown density structure of the HII 
region creates some uncertainty. If there are
significant variations in optical depth with frequency then
it is possible that lines at different frequencies could
be generated in different locations within the HII region.

Aside from the systematic effects, the uncertainty in the average 
electron density is given
by the standard propagation of errors, assuming the errors in the
widths are independent. 
Uncertainties in the electron densities are listed in table \ref{electrondensities} 
and for the line center velocities
and widths in table \ref{CMrrl}.

\subsection{Electron densities from the radio continuum}

At the high densities in our HII regions, the
centimeter continuum emission is optically thick to free-free emission. 
This can be
determined from the formula 
\citep[eqn. A.1b,][]{MezgerHenderson1967}
or \citep[eqn. 10,][]{Keto2003},
\begin{equation}\label{eq:opticalDepth}
\tau_\nu = 8.235\times 10^{-2} \bigg({{T_e}\over{{\rm K}}}   \bigg) ^{-1.35}
\bigg({{\nu}\over{{\rm GHz}}}   \bigg) ^{-2.1}
\bigg({{\rm n_e^2L}\over{{\rm pc\ cm}{^{-6}}}}   \bigg)
 \end{equation}
where $T_e$ is the electron temperature and $n_e^2L$ is
the electron density squared times the path length.
For a characteristic size of 0.01 pc, an HII region has an optical depth
of unity at 22 GHz (1.3 cm) if the electron density is $4\times 10^5$ cm$^{-3}$. 
This means
that the formulae for optically thin emission
\citep[e.g.][]{MezgerHenderson1967} are not useful
for the HII regions in our sample,
and probably not for small, high-density HII regions in general.

If the gas is optically thick to free-free emission, and the
electron density is estimated from the radio continuum by incorrectly
assuming optically thin emission, then the derived electron density
is just that density required to make the optical depth unity.
This can be understood by a simple example.
The radio free-free
emission through a slab of uniform gas density and temperature is,
\begin{equation}\label{eq:exactEmission}
S_\nu =  2k T_e (\nu^2/c^2)
\big[1 - e^{-\tau_\nu}\big] ({\rm Jy/Sr})
\end{equation}
The optically thin approximation amounts to replacing
$[1-\exp(-\tau_\nu)]$ by $\tau_\nu$ so that,
\begin{equation}\label{eq:thinEmission}
S_\nu \approx  2k T_e (\nu^2/c^2)\tau_\nu
\end{equation}
whereas, in the optically thick limit,
\begin{equation}\label{eq:thickEmission}
S_\nu \approx  2k T_e (\nu^2/c^2)
\end{equation} 
Thus if the HII region is optically thick (equation \ref{eq:thickEmission}) 
but $\tau_\nu$ is derived assuming equation \ref{eq:thinEmission}, then the
derived value of the optical depth is 1, 
and the electron density derived from equation \ref{eq:opticalDepth}
is the density required to make $\tau_\nu=1$. 
Thus at 22 GHz (1.3 cm),
the electron density derived for an HII region with a size of 0.01 pc is never
be greater than about $10^5$ cm$^{-3}$ if the optically thin formulae are used --
regardless of the true density.  
This underestimate also affects
quantities derived from the electron density such as 
the number of ionizing photons required to
balance recombinations which in turn is often used to derive the
spectral types of the exciting stars.

Formulas for spherical HII regions that have
a density gradient such that the central part of the HII region is optically thick
are derived in \citet{Keto2003}. However, these formulae require a 
specification of the size of the optically thick region.

\subsection{The correlation of line widths, velocities, and frequencies}

As shown in table \ref{CMrrl}, the line widths decrease with
increasing frequency in all cases.  This suggests the increasing
importance of pressure broadening at lower frequencies.  For sources
with more than one centimeter RRL observed, the electron densities
increase with the increasing frequency of the centimeter line used in
the analysis. The one exception is the electron density calculated
from the H66$\alpha$ line in NGC7538-IRS1. This line is very much
wider and out of character with the other measurements in our sample.
Aside from this one exception, the correlation of density with
observing frequency indicates that the higher frequency lines are
associated with higher density gas.  The line center velocities also
increase (redshift) with frequency, except for the lines of
NGC7538-IRS1.

The combination of the two correlations, line center
velocity with frequency and line width with frequency was first noticed
in W3(OH) by \citet{BerulisErshov1983}.  \citet{WelchMarr1987} and
\citet{Keto1995} interpreted the dual correlation as evidence for an 
accelerating flow within a density gradient \citep{Guilloteau1983}.  
A full explanation is given in \citet{Keto1995} (see also 
\citet{BrocklehurstSeaton1972})and can be summarized as follows:
The peak intensity of RRLs is proportional to
the density squared and inversely proportional to the line width. 
Thus the peak intensity of optically thin
high frequency lines with negligible pressure 
broadening is proportional to the density squared.
The peak intensity of lower frequency lines is proportional
to the first power of the density since the pressure broadened
line width is itself proportional to density.
Thus in spatially unresolved observations of gas with a density gradient,
the higher frequency RRLs are dominated by emission from the higher density
gas while the lower frequency RRLs include contributions from both high and
low density gas. An observed difference in velocity between the high
and low frequency lines thus indicates a difference in velocity between the
high and low density gas.

The correlation of velocities and width with frequency therefore indicates
that in these HII regions there is both a flow and a density gradient. The two
are naturally related by the conservation of mass and the geometry of the
flow. In general, if the flow is either outward and diverging or inward and
converging then conservation of mass requires that there be a density gradient
except in the unique circumstance that
the variation of the flow speed exactly cancels the geometric
divergence or convergence.

\section{Continuum Spectral Indices}

We determined the spectral energy distrubutions (SEDs) of the HII
regions in our sample using our own millimeter and centimeter
observations and data collected from the literature. 
We selected only those data with
angular resolutions close to 1$^{\prime\prime}$ specifically
excluding single dish observations. 
Most of the data for NGC7538-IRS1 and G45.07+0.14 are from 
\citet[][table 1]{Pratap1992} and \citet[][table 3]{Garay1986}, 
respectively. The rest of the observations are 
listed in table \ref{fluxdata}.
Most of these data 
do not include estimated uncertainties. However, the calibration
of radio data is fairly standard. Many of the reported observations
were made at the VLA that typically calibrated flux densities 
to better than 20\% at 1.3 cm at the dates of the observations
and much better at longer wavelengths.
As noted in \S \ref{observations}, we estimated our SMA 
observations to be calibrated to better than 39\%. 

The spectral energy distributions are shown in figures 6 through 10.
Along with each observed SED, shown as crosses, 
the figures show two model SEDs as solid lines.  One model SED
is based on a single HII region with a power-law
density gradient. The continuum flux density for this model
is calculated as in \citet{Keto2003}. 
The radio free-free 
emission of a spherical HII region seen as an unresolved source is,
\begin{equation}
S_\nu = 4\pi k T_e \nu^2/c^2  
\int^{\theta_0}_0 \theta\big[1 - e^{-\tau_\nu}\big]d\theta
\end{equation}
Here, $\theta$ is an angular coordinate on the plane of the
sky, 
$\theta = \sqrt{x^2 + y^2}/D$
with coordinate axes $x$ and $y$ are centered on the HII region,
$\theta_0 = R_0/D$ where 
$R_0$ is
the radius of the HII region, and $D$
the  distance to the source.
An approximation to the free-free optical depth is,
\citet[][equation A.1a]{MezgerHenderson1967}, 
\begin{equation}
\tau_\nu = 8.235\times 10^{-2} \bigg( {{T_e}\over{^\circ K}} \bigg) ^{-1.35}
        \bigg( {{\nu}\over{GHz}} \bigg)^{-2.1}
        \bigg( {{EM}\over{{\rm pc}\ {\rm cm}^{-6} }} \bigg)
\end{equation}
where $T_e$ is the electron temperature, and $EM$ is the emission measure, $n_e^2L$.

The model parameters for all sources
are listed in table \ref{ModelSED}.  The other model is an SED based
on constant density HII regions.  The figures for NGC7538-IRS1 and W51e2 
show the SEDs from multi-component models suggested
by \citet{Pratap1992} and \citet{Rudolph1990}. These are both based on
constant density HII regions plus dust.  The dust emission enters as
\citep[][ equation 1]{Pratap1992},
\begin{equation}\label{eq:dust}
S_\nu({\rm Jy}) = 52.36 {M}_{\odot} 
\bigg(  {{0.2}\over{\lambda_{mm}}} \bigg)^{\beta+3}
\bigg[ \exp \bigg(  {{14.4}\over{\lambda_{mm}T_d}} \bigg) -1\bigg]^{-1}D^{-2}_{kpc}
\end{equation}
where $S_\nu$ is the dust emission in Jy, $M$ is the mass of the
dusty gas in M$_\odot$, $\beta$ is the spectral index of the dust 
emissivity, $T_d$ is the
dust temperature in Kelvin, and $D$ is the distance in kpc.

The multi-component model for NGC7538-IRS1 \citep{Pratap1992} nicely
illustrates how a density gradient within an HII region can stretch
the transition between optically thick and thin frequencies
resulting in an SED with an intermediate spectral index. In this
model there are two concentric spherical regions each of uniform but different
densities. 
The smaller, higher density component produces an
emission curve that is shifted upward and to the right of the curve
for the larger, lower density component (figure 6). The emission from
the two components combines to produce an SED of intermediate slope.
An HII region with a continuous
density gradient can be considered as the logical extension of this
two component model to many nested smaller and denser HII
regions.  \citet{Lugo2004} also model the NGC 7538-IRS1 HII region with
a density gradient that they suggest is due to a wind off a
photo-evaporating disk.

The SED modeling combined with the recombination line measurements
shows that dust is not a significant contributor to the continuum
emission measured at high angular resolution from these sources.  (See
also \citet{Kurtz2005}, \citet{Depree1998}, and \citet{Depree2000}.)
Consider W51e2 as an example.  As seen in figure 7, the HII region can
be modeled with uniform, low-density ionized gas, with the 147 and
230~GHz emission arising from dust \citep{Rudolph1990}.  Figure 7
shows that with this approach, the free-free continuum emission at 230
GHz would be about a factor of 10 below the observed emission.
However, the line-to-continuum ratio should be approximately one at
this frequency \citep{GordonS2002}. Hence, if the free-free continuum
were this low, the H30$\alpha$ recombination line would also have a
much lower flux density --- so low that the line would not be detected
in our observations.  Stated another way, the recombination line
emission cannot be 10 times greater than the free-free continuum
emission at this frequency.  Therefore, most of the continuum emission
observed by the interferometer at 231.901 GHz must be free-free, hence
the electron density must be higher, and a density gradient is
required.

The high frequency emission observed in large beams ($20-30^{\prime\prime}$)
around these bright HII regions
certainly arises from dust. Again take W51e2 as an example.
\citet{Jaffe1984} report a flux density
of 1200 Jy~beam$^{-1}$ at 750 GHz and 40$^{\prime\prime}$ angular
resolution. We assume, as in the model of \citet{Rudolph1990}, that
this emission is from cool dust in the large-scale molecular cloud
surrounding the HII region.  
Because the dust is cool (50 - 100 K), a large mass of dust is required
(4500 M$_\odot$, \citet{Rudolph1990}) to produce the observed 750 GHz flux
density.
While the emission from this dust is also sufficient to produce the 
90 and 230 GHz flux density 
that is observed by interferometers,  the cool dust
is not the source of this emission because it is not possible that this
large mass of cool dusty gas is within the spatial scale defined by the
arc second angular resolution of the interferometers.
However, the ionized gas is sufficiently hot
($10^4$ K) that only a small mass (0.01 M$_\odot$), consistent with
the mass of the HII regions, is required to produce the 90 - 230 GHz
flux density. Similar considerations apply to the multi-component models of
other HII regions, for example, the model discussed earlier for
NGC7538-IRS1 \citep{Pratap1992}.

On a spatial scale similar to that of HCHII
and UCHII regions, continuum emission from dust has been observed 
by the SMA in the nearby (1.7 kpc) massive star forming region NGC6334 by
\citet{Hunter2006}.  The dusty clumps of arcsecond scale in NGC6334
have continuum emission on the order of 1~Jy.  Scaled to the
distance of W51, the dust in such clumps would be a minor contributor
(60 mJy) to the continuum emission.  Thus, high angular resolution
observations of dusty clumps in NGC6334 are consistent with the
interpretation of negligible dust contribution to the emission
observed at high angular resolution in more distant bright
HII regions.

\section{Discussion}\label{discussion}

The observations suggest that pressure broadening of the centimeter
RRLs is important in small HII regions. This contradicts the
interpretation of some previous studies. In particular, on the basis
of electron densities calculated by assuming constant gas density and
optically thin emission, \citet{Gaume1995a}, \citet{Depree1997}, and
\citet{Depree2005} have suggested that the electron densities in
NGC7538-IRS1, W49, and SgrB2 are too low to cause significant pressure
broadening in the H66$\alpha$ line.  However, the RRL observations
suggest that the electron density in NGC7538-IRS1 is two orders of
magnitude greater than the $10^5$ cm$^{-3}$ derived by
\citet{Gaume1995a} and sufficient to cause significant pressure
broadening.  \citet{JaffeMP1999} report observations of millimeter RRL of
several sources, including G10.6-0.4 and NGC7538-IRS1.  They concluded
that the RRL widths are not due to pressure broadening because the
observed widths of different recombination lines do not scale as the
fourth power of the frequency.   
Their conclusion was based on a comparison of observations with
different angular resolutions. It has been demonstrated that low
angular resolution observations may suffer confusion from
more-extended, lower-density gas \citep{Simpson1973a, Simpson1973b,
GS1974, Smirnov1984, GordonS2002}.  We concur with
\citet{Gaume1995a, Depree1997, Depree2005, JaffeMP1999} 
that broadening due to gas motions is a significant contribution to
the observed line widths, but we find that pressure broadening is also significant,
particularly for the lower frequency lines.

The intermediate SEDs of small HII regions
imply a density gradient as would arise from a divergence or convergence in
the flow of ionized gas in an HII region. 
Other interpretations are possible. 
\citet{Franco2000}, \citet{KurtzFranco2002} and \citet{KimKoo2002} 
suggested that the
various densities are due to the hierarchical structure of the
molecular cloud prior to its ionization by a massive star.
The intermediate SEDs can also be created by density variations 
other than gradients.
\citet{Ignace2004} suggested that
an HII region is actually an ensemble of differently-sized clumps,
each of uniform but different density. One problem with these models of
pre-existing structure or clumps is that their survival time 
is approximately their sound crossing time.
This timescale seems impossibly short if the high density clumps are
substructures within an HCHII  ($t_c \leq 1000$ yrs for $R < 0.01$ pc).

The interpretation of density gradients and resulting high electron
densities in small HII regions also contradicts the interpretation of
significant dust emission in a number of other studies.  For example,
\citet{Rudolph1990, Pratap1992, Testi2000, Schilke1990, Beuther2004,
  Beuther2006} model the small HII regions, W51e2, NGC7538-IRS1,
G9.62+0.19-F, and Orion-KL-SMA1 respectively, with low and constant
density gas and assume that dust makes up the emission at high
frequencies.  \citet{Testi2000} also provide an alternative model with
a stellar wind rather than dust to supply the high frequency emission.
These models with low and constant density have difficulty in
explaining the broad RRL line widths. If the density is low, then the
widths must be due to dynamical broadening created by a supersonic
flow.  However, a constant density is not generally consistent with
supersonic flows. In general, a strong density 
gradient is required to create the
strong pressure gradient needed to drive the flow. 

\section{Conclusions}

1. The radio recombination line widths increase with decreasing 
   frequency in all cases. This indicates the greater importance 
   of pressure broadening at lower frequencies.

2. The electron densities calculated by comparison of 
   centimeter recombination lines with the H30$\alpha$ millimeter line
   increase with increasing frequency of the centimeter line.
   This suggests that the higher frequency lines are generated in 
   denser gas and suggest that density gradients exist in small
   HII regions.

3. Supersonic line widths are observed in the H30$\alpha$ RRL.
   This line is at a frequency high enough that pressure
   broadening is negligible provided that the density
   of the emitting gas is less than $10^8$ cm$^{-3}$. The large
   line widths suggest the presence of supersonic motions
   within the HII regions.

3. The electron densities in HCHII and UCHII regions are generally
   high enough that the gas is optically thick at centimeter
   wavelengths. This means that electron densities cannot be calculated
   from radio continuum observations 
   using formulae that assume optically thin emission.
   Instead formulae appropriate for partially
   optically thick emission must be used.

4. At high frequencies ($\geq 100$ GHz) and high angular resolution
   ($\sim 1^{\prime\prime}$) the observed continuum emission is mostly
   free-free emission and not from dust.  Comparisons based on high
   frequency observations with different beam sizes can lead to erroneous
   conclusions. High frequency emission at low angular resolution
   ($\geq 10^{\prime\prime}$) is probably from dust, but associated with the
   large-scale, overlying molecular cloud.

S. Kurtz acknowledges support from UNAM, DGAPA project IN106107.

\clearpage

\begin{deluxetable}{lrrccccclc}
\tabletypesize{\scriptsize}
\tablecolumns{10}
\tablewidth{0pt}
\tablecaption{Calibration of Millimeter Observations \label{millimetercalibration}}
\tablehead{
&   \multicolumn{2}{c}{Flux} &   
\colhead{} & \multicolumn{2}{c}{Bandpass}  
& \colhead{} & \multicolumn{2}{c}{Gain} \\
\cline{3-4} \cline{6-7} \cline{9-10}\\
\colhead{Source}  & Band & \colhead{Calibrator}   & \colhead{Flux}  &  & \colhead{Calibrator} & \colhead{Flux} & &
 \colhead{Calibrator}   & \colhead{Flux}     \\
 & & & \colhead{Jy}   & & & \colhead{Jy}  & & & \colhead{Jy} 
} 
\startdata
G10.60 -0.4       & SMA 	& Ceres	& 0.7 	&& 3C454.3	& 16.8	&& PKS 1911-201	& 1.4 \\
G28.20 - 0.04 N  & SMA	& \nodata& \nodata&& 3C454.3	& 14.3	&& PKS 1911-201	& 1.8 \\
G45.07 + 00.14    & SMA	& \nodata& \nodata&& 3C454.3	& 4.6	&& PKS 2025+337	& 0.5 \\
W51e2        	& SMA	& Ceres	& 0.7 	&& 3C454.3	& 16.0	&& PKS 2025+337	& 0.8 \\
NGC7538-IRS1 	& SMA	& Ceres	& 0.5 	&& 3C454.3	& 11.9	&& PKS 2202+422	& 3.0 \\
\enddata
\tablecomments{Flux densities derived assuming a 
flux density for Uranus of 36.6 Jy. }
\end{deluxetable}


\begin{deluxetable}{lrrccccccl}
\tabletypesize{\scriptsize}
\tablecaption{VLA Observations  \label{VLAobs}}
\tablewidth{0pt}
\tablehead{
\colhead{Source} &  \colhead{RA} & \colhead{Dec} & \colhead{Line} & \colhead{Frequency} & \colhead{Spectral} &
\colhead{Band} &
\colhead{Angular} & \colhead{Cont.\tablenotemark{a}} & \colhead{Date} \\
&  &  & & & \colhead{Resolution} & \colhead{width} & \colhead{Resolution} & \colhead{rms} & \colhead{Observed} \\
& & & &\colhead{GHz} & \colhead{kms$^{-1}$} & Chan. & \colhead{arc second} & \colhead{mJy} 
}
\startdata
G10.6 -0.4   & 18 10 28.7  & -19 55 49 & H92$\alpha$ & 8.309383 & 1.76  & 64 & $1.5 \times 1.0$ & 7   & 1989-03-17\\
 & & & H66$\alpha$\tablenotemark{b} & 22.36417 & 2.62  & 64 & $1.7 \times 1.0$ & 9   & 1988-05-23\\
 & & & H53$\alpha$ & 42.95197 & 1.36  & 64 & $1.5 \times 1.3$ & 3.0 & 2003-02-03\\
G28.20 - 0.04 N & 18 42 58.17  & -04 13 57.0 & H53$\alpha$ & 42.95197 & 1.36  & 64 & $0.6 \times 0.4$ & 1.2 & 2005-06-24\\
& & & H53$\alpha$ & 42.95197 & 5.45  & 32 & $0.7 \times 0.5$ & 3.0 & 2005-06-28\\
G45.07+0.14 N& 19 13 22.069 &  10 50 52.5 
                     & H76$\alpha$\tablenotemark{c}   & 14.69000 & 15.95 & 16 & $0.6 \times 0.6$ & 10  & 1984-01-07\\
W51e2    & 19 23 43.913 &  14 30 14.7             & H66$\alpha$ & 22.36417 & 2.62  & 64 & $0.3 \times 0.2$ & 1.6 & 2003-02-01\\
& & & H53$\alpha$ & 42.95197 & 1.36  & 64 & $0.5 \times 0.4$ & 0.9 & 2003-02-01\\
NGC7538-IRS1  & 23 13 45.37  &  61 28 10.4       & H53$\alpha$ & 42.95197 & 5.45  & 31 & $0.2 \times 0.2$ & 0.9 & 2005-04-17\\
\enddata
\tablenotetext{a}{Bandwidth = channels $\times$ spectral resolution}
\tablenotetext{b}{Data from the VLA archives. The original observation is reported in \citet{Keto2002a}.}
\tablenotetext{c}{Data from the VLA archives. The original observation is reported in \citet{Garay1986}.}
\end{deluxetable}

\begin{deluxetable}{lccccccccc}
\tabletypesize{\scriptsize}
\tablecolumns{10}
\tablewidth{0pt}
\tablecaption{Calibration of Centimeter Observations \label{centimetercalibration}}
\tablehead{
&   \multicolumn{2}{c}{Flux} &   
\colhead{} & \multicolumn{2}{c}{Bandpass}  
& \colhead{} & \multicolumn{2}{c}{Gain} \\
\cline{3-4} \cline{6-7} \cline{9-10}\\
\colhead{Source}  & Band & \colhead{Calibrator}   & \colhead{Flux}  &  & \colhead{Calibrator} & \colhead{Flux} & &
 \colhead{Calibrator}   & \colhead{Flux}     \\
 & & & \colhead{Jy}   & & & \colhead{Jy}  & & & \colhead{Jy} 
} 
\startdata
G28.20 - 0.04 N 	& Q	& 3C286	& 1.5 	&& 3C273  	& 24.2	&& PKS 1851+005	& 1.9 \\
G45.07 + 00.14 N & U	& 3C286	& 3.5 	&& 3C454.3	& 7.5	&& PKS 1923+210	& 2.1 \\
W51e2        		& Q	& 3C286	& 1.5 	&& 3C273  	& 10.1	&& PKS 1923+210	& 2.6 \\
W51e2        		& K	& 3C286	& 2.5 	&& 3C273  	& 19.5	&& PKS 1923+210	& 3.1 \\
NGC7538-IRS1 		& Q	& 3C48	& 0.5 	&& 3C454.3	& 15.6	&& PKS 2250+558	& 0.8 \\
\enddata
\end{deluxetable}

\begin{deluxetable}{lccccccccc}
\tabletypesize{\scriptsize}
\tablecaption{Recombination Line data \label{CMrrl}}
\tablewidth{0pt}
\tablehead{
\colhead{Source}  & \colhead{Continuum} & 
\colhead{Integrated} & \colhead{Quantum} &
\colhead{Line} & \colhead{$1 \sigma$} & \colhead{Line} & \colhead{$1 \sigma$} & \colhead{Line} & \colhead{$1 \sigma$} \\
&   \colhead{Peak} & 
\colhead{Intensity} & \colhead{Number} &
\colhead{Peak} & & \colhead{Velocity\tablenotemark{a}} & & \colhead{Width\tablenotemark{b}} & \\
&   \colhead{Jy/beam} & \colhead{Jy} & &
\colhead{mJy/beam} & \colhead{mJy/beam} & \colhead{kms$^{-1}$} &  \colhead{kms$^{-1}$} & 
\colhead{kms$^{-1}$} &\colhead{kms$^{-1}$}
}
\startdata 
G10.6-0.4  & 0.017 & 3.14 & 92& 8.1 & 0.2  & 0.6 & 0.5  & 37.5 & 1.2\\
& 0.04 \tablenotemark{c}  & 4.57 & 66& 188.0 & 5.0 & 2.1 & 0.005  & 35.1 & 0.005 \\
&  2.1  & 6.0  & 53& 219.0 & 4.9 & 2.2 & 0.005  & 31.8 & 0.01 \\
&  1.17  & 2.41 & 30 & 1040.0 & 28.5 & 3.4 & 0.3 & 24.5 & 0.8 \\
G28.20-0.04 N& 0.4  & 1.1  & 53& 99.0 & 3.6 & 88.7 & 0.6 & 33.4 & 1.5 \\
 & 0.56  & 0.72 & 30 &930.0 & 22.0  & 92.5 & 0.2 & 20.9 & 0.6 \\
G45.07+0.14 & 0.37 & 0.77 & 76& 21.0 & 2.5 & 45.7 & 3.4 & 56.9 & 8.1\\
& 0.08  & 0.17 & 30 & 92.0 & 10.2 & 49.4 & 1.8 &33.2 & 4.2 \\
W51e2              & 0.15 & 0.30 & 66& 14.2 & 0.4 & 53.8 & 0.7 & 50.9 & 1.6 \\
 & 0.17 & 2.3  & 53& 42.0 & 1.1 & 59.5 & 0.3 & 32.5 & 1.2 \\
 & 1.67  & 3.66 & 30 &870.0 & 51.0 & 59.7 & 0.8 & 26.8 & 1.9\\
NGC7538-IRS1	  & 0.2  & 0.7  & 53& 23.3 & 1.1 & -41.8 & 1.0 & 61.0 & 4.1 \\
 & 2.56  & 2.79 & 30 & 4450.0 & 58.0 & -60.5 & 0.4 & 57.3 & 0.8 \\
\enddata
\tablenotetext{a}{ {Line} center velocity from Gaussian fit.}
\tablenotetext{b}{FWHM from Gaussian fit.}
\tablenotetext{c}{Keto 2002}

\end{deluxetable}


\begin{deluxetable}{lcccccc}
\tabletypesize{\scriptsize}
\tablecaption{Electron Densities from line widths \label{electrondensities}}
\tablewidth{0pt}
\tablehead{
\colhead{Source} & \colhead{Quantum Number} & \colhead{Line Width} &
\colhead{$1 \sigma$} & \colhead{$n_e$} & \colhead{$1 \sigma$}  \\
&   &    \colhead{kms$^{-1}$} & \colhead{kms$^{-1}$} & \colhead{cm$^{-3}$} & \colhead{cm$^{-3}$} }
\startdata
G10.6-0.4       & 92    & 37.5                  & 1.2   & $9.1\times 10^4$ & $7.3\times 10^3$ \\
                & 66    & 35.1\tablenotemark{a} & 0.005 & $7.8\times 10^5$ & $3.2\times 10^2$ \\
                & 53    & 31.8                  & 0.01  & $2.5\times 10^6$ & $3.2\times 10^3$ \\
G28.20-0.04 N   & 92    & 74.4\tablenotemark{b} & 2.6   & $2.9\times 10^5$ & $1.2\times 10^4$   \\
                & 53    & 33.4                  & 1.5   & $4.1\times 10^6$ & $4.3\times 10^5$   \\
G45.07+0.14     & 76    & 56.9\tablenotemark{d} & 8.1   & $6.1\times 10^5$ & $1.8\times 10^5$   \\
                & 66    & 42.3\tablenotemark{e} & 2.3   & $6.9\times 10^5$ & $1.6\times 10^5$   \\
W51e2           & 66    & 50.9                  & 1.6   & $1.6\times 10^6$ & $9.0\times 10^4$   \\
                & 53    & 32.5                  & 1.2   & $2.0\times 10^6$ & $4.0\times 10^5$  \\
NGC7538-IRS1    & 66    & 180\tablenotemark{c}  & ...   & $1.0\times 10^7$ & ...        \\
                & 53    & 61.0                  & 4.1   & $1.4\times 10^6$ & $1.5\times 10^6$   \\
\enddata
\tablecomments{Electron density, $n_e$,
required to produce the observed increase in the width of the
cm RRL over the mm RRL by pressure broadening. }
\tablenotetext{a}{Keto 2002}
\tablenotetext{b}{Sewilo et al 2004}
\tablenotetext{c}{Gaume et al 1995a}
\tablenotetext{d}{Garay et al 1986. The line width measured in our analysis of this data is larger than reported
in the original reference (48.1 kms$^{-1}$).}
\tablenotetext{e}{Garay et al 1985}
\end{deluxetable}

\begin{deluxetable}{lcccl}
\tabletypesize{\scriptsize}
\tablecaption{Integrated Intensities from Previous Observations\label{fluxdata}}
\tablewidth{0pt}
\tablehead{
\colhead{Source} & \colhead{Frequency} & \colhead{Flux} & \colhead{Beam} & \colhead{Reference} \\
		&  \colhead{GHz} 	& \colhead{Jy} 	& \colhead{Arc seconds} }
\startdata
G10.6-0.4	& 23 	& 2.4 		& $0.1 $ 	  & \citet{SollinsHo2005} \\
	& 22.3	& 4.1 		& $0.2\times 0.1$ & \citet{KetoWood2006} \\
	& 15 	& 2.8 		& $0.3 $	  & \citet{HoHaschick1981} \\
	& 15 	& 2.6 		& $0.3\times0.2 $ & \citet{TurnerMatthews1984} \\
	& 8.3	& 3.1 		& $1.5\times 1.0$ & this paper \\
	& 5 	& 1.7 		& $1 $ 		  & \citet{HoHaschick1981} \\
G28.20-0.04 N	& 23.7  & 0.98 		& $0.3\times 0.2$ & \citet{Sollins2005b}	\\
	& 15    & 0.54 		& $0.5$		  & \citet{Kurtz1994}	\\
	& 8.4   & 0.30 		& $0.9$		  & \citet{Kurtz1994}	\\
	& 8.3   & 0.30 		& $4.9\times 2.7$ & \citet{Sewilo2004}	\\
G45.07+0.14	& 345   & 2.71 		& $3.0\times 4.0$ & \citet{Su2004}	\\
	&  98   & 1.04 		& $2.6\times 2.4$ & \citet{Hunter1997}	\\
W51e2			& 147	& 3.7  		& $2.0$ 	   & \citet{ZhangHoOhashi1998}\\
		& 88.2  & 1.23 		& $6 \times 5$     & \citet{Rudolph1990}\\
			& 23.7  & 0.35 		& $5 \times 4$     & \citet{Rudolph1990}\\
			& 14.7  & 0.26 		& $4.1 \times 3.9$ & \citet{Garay1985}\\
			& 8.3   & 0.07 		& $2.4 $ 	  & \citet{Mehringer1994}\\
			& 5.0   & $\leq 0.03$ 	& $4.1 $ 	  & \citet{Mehringer1994}\\
			& 5.0   & $0.025$ 	& $0.5 $ 	  & \citet{Rudolph1990}\\
\enddata
\end{deluxetable}

\begin{deluxetable}{lccccccc}
\tabletypesize{\scriptsize}
\tablecolumns{8}
\tablewidth{0pt}
\tablecaption{Model HII regions for continuum SEDs\label{ModelSED}}
\tablehead{
&  &  \multicolumn{2}{c}{Constant Density} &   \colhead{} &
\multicolumn{3}{c}{Density Gradient} \\
\cline{3-4} \cline{6-8} \\
\colhead{Source} & \colhead{Distance}   & \colhead{radius}    & \colhead{density} & \colhead{} &
 \colhead{radius}   & \colhead{density}    & \colhead{exponent} \\
 & \colhead{kpc}   & \colhead{pc}    & \colhead{cm$^{-3}$} & \colhead{} &
 \colhead{pc}   & \colhead{cm$^{-3}$}   } 
\startdata
G010.6 - 0.4 & 6.0	& 0.05	& $6.0\times 10^4$	&& 0.05	& $3.5\times 10^4$ & -1.4	\\
G28.20 - 0.04 N & 9.1	& 0.03	& $9.0\times 10^4$	&& 0.043	& $1.0\times 10^4$ & -2.5\\
G045.07 + 00.14 & 9.7	& 0.025	& $1.6\times 10^5$	&& 0.05	& $1.5\times 10^4$ & -1.7	\\
W51e2\tablenotemark{a} 	& 7.0	& 0.01	& $3.0\times 10^5$	&& 0.01	& $1.0\times 10^5$ & -2.5	\\
NGC7538-IRS1\tablenotemark{b} 	& 3.5	& 0.02	& $2.7\times 10^6$	&& 0.08	& $2.7\times 10^6$ & -2.0 \\
NGC7538-IRS1\tablenotemark{c} 	& \nodata& 0.08	& $5.0\times 10^4$	&& \nodata	& \nodata & \nodata \\
\enddata
\tablenotetext{a}{SED also includes a contribution from
dust modeled as 4500 M$_\odot$ of cold, dusty gas with a 
temperature of 100 K and a dust emissivity, $\beta=1.5$ from the 
model of \citet{Rudolph1990}.}
\tablenotetext{b}{Constant density model with 3 components, 
2 HII regions plus dust  from the model of \citet{Pratap1992}.
The first HII component SED also includes a contribution from
dust modeled as 53 M$_\odot$ of cold, dusty gas with a 
temperature of 50 K and a dust emissivity, $\beta=1.0$}
\tablenotetext{c}{Second component of constant density model. }
\end{deluxetable}

\clearpage

\begin{figure}
\includegraphics[angle=90,width=4.truein]{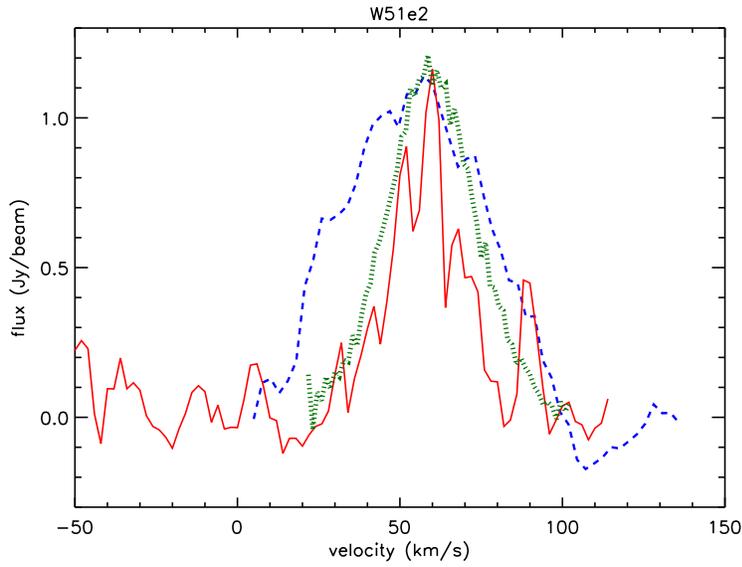}
\caption{ H66$\alpha$ (dashed blue line), H53$\alpha$ (dotted green line),
and H30$\alpha$ (solid red line) in
W51e2. The flux densities of the H66$\alpha$ line and the H53$\alpha$ lines
have been multiplied by 80 and 27, respectively.}
\label{fig:w51spectra}
\end{figure}

\begin{figure}
\includegraphics[angle=90,width=4.truein]{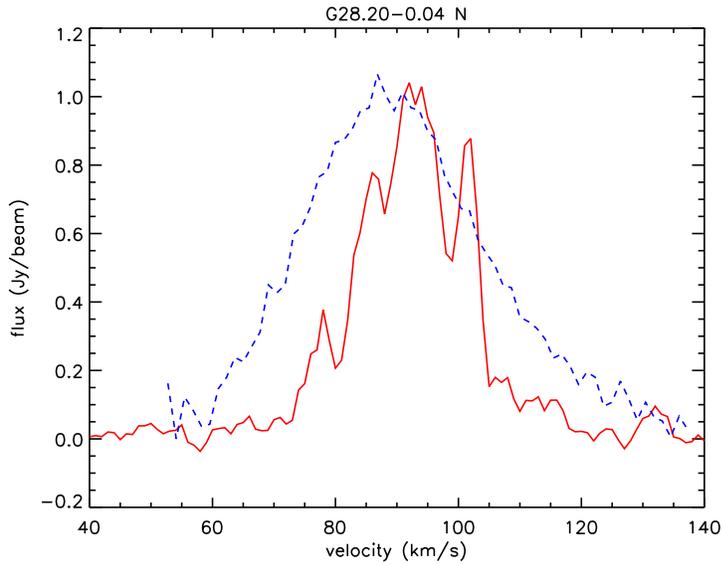}
\caption{H53$\alpha$ (dashed blue line) and H30$\alpha$ (solid red line) in
G28.20-0.04N. 
The flux density of the H53$\alpha$ line has been multiplied by 10.}
\label{fig:g28nspectra}
\end{figure}

\begin{figure}
\includegraphics[angle=90,width=4.truein]{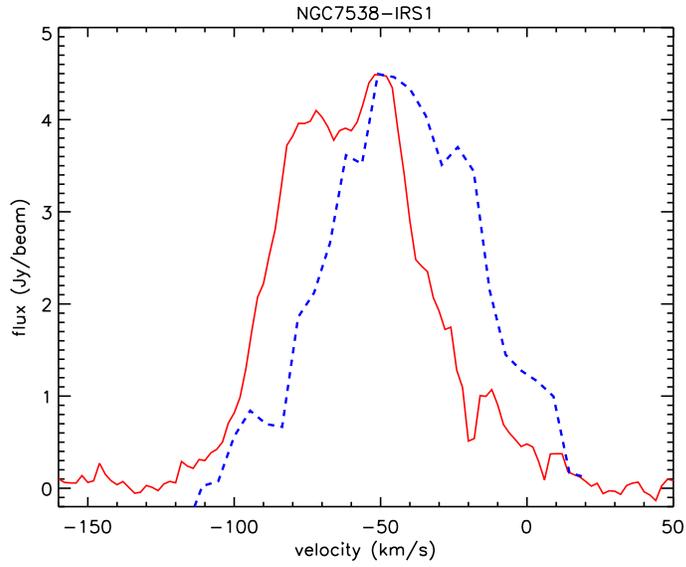}
\caption{ H30$\alpha$ (solid red line) and H66$\alpha$ (dashed blue line)
in NGC7538-IRS1. The flux density of the H66$\alpha$ has been multiplied by
190.}
\label{fig:n7538spectra}
\end{figure}

\begin{figure}
\includegraphics[angle=90,width=4.truein]{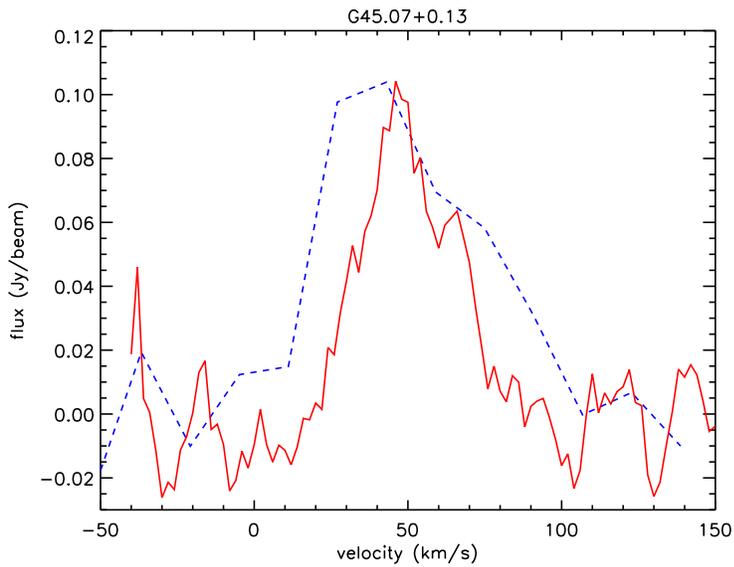}
\caption{ H76$\alpha$ (dashed blue line) and H30$\alpha$ (solid red line) in
G45.07+0.13. The flux density of the H76$\alpha$ line has been multiplied by 5.}
\label{fig:g45spectra}
\end{figure}

\begin{figure}
\includegraphics[angle=90,width=4.truein]{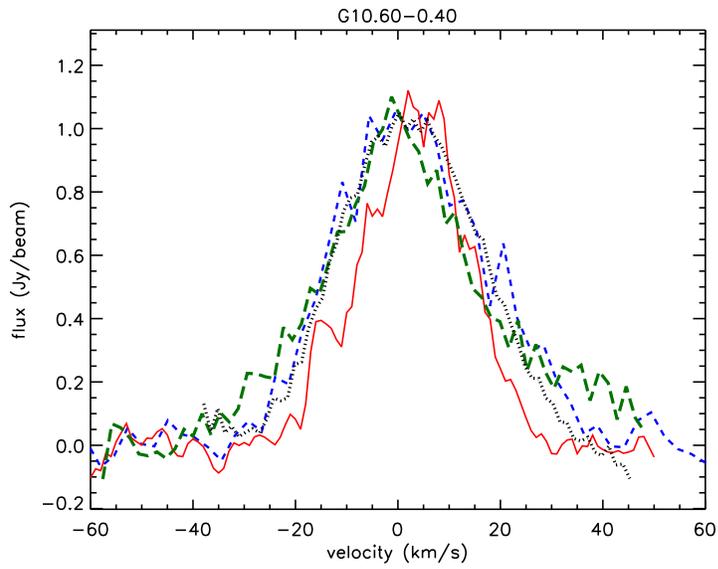}
\caption{ H92$\alpha$ 
(long-dashed green line), 
H66$\alpha$ 
(short-dashed blue line), 
H53$\alpha$ (dotted black line),
and H30$\alpha$ (solid red line) in G10.6-0.04. The flux densities 
of the H92$\alpha$,
H66$\alpha$, and H53$\alpha$ lines have been multiplied by 
115, 5.5, and 4.8 respectively.}
\label{fig:g106spectra}
\end{figure}

\begin{figure}
\includegraphics[angle=90,width=4.truein]{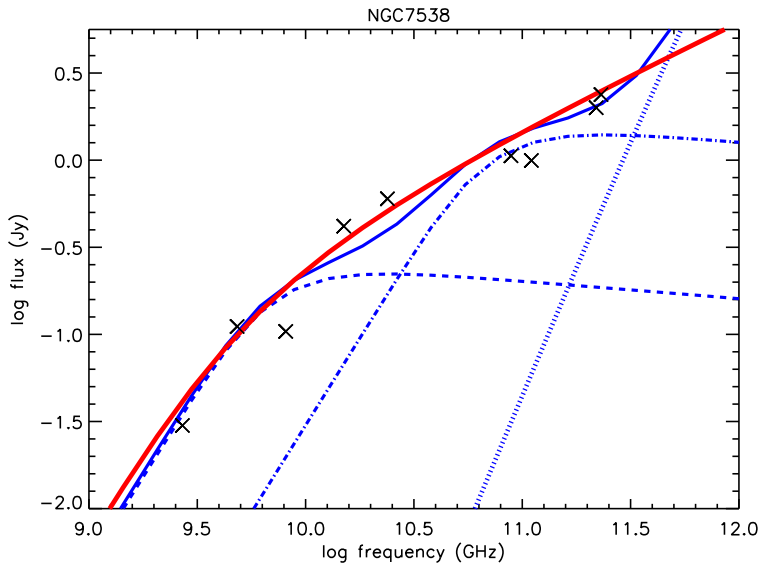}
\caption{SED NGC 7538 IRS1. Data are from table 1 of \citet{Pratap1992} with
the addition of our 230 GHz SMA observation. The blue lines show a 3 component model
proposed by \citet{Pratap1992} consisting of two HII regions plus dust 
(dotted line). 
The dash-dotted line (of the smaller, higher density component)
is shifted upward and to the right of the dashed line (of the larger, lower
density component. The wavy solid blue line shows the sum of the three
components.
The red solid line that crosses smoothly through the waves of
the three component model shows the SED of a single HII 
region with a power-law density gradient.  See \S \ref{derivation}
and table 7.}
\label{fig:sedn7538}
\end{figure}

\begin{figure}
\includegraphics[angle=90,width=4.truein]{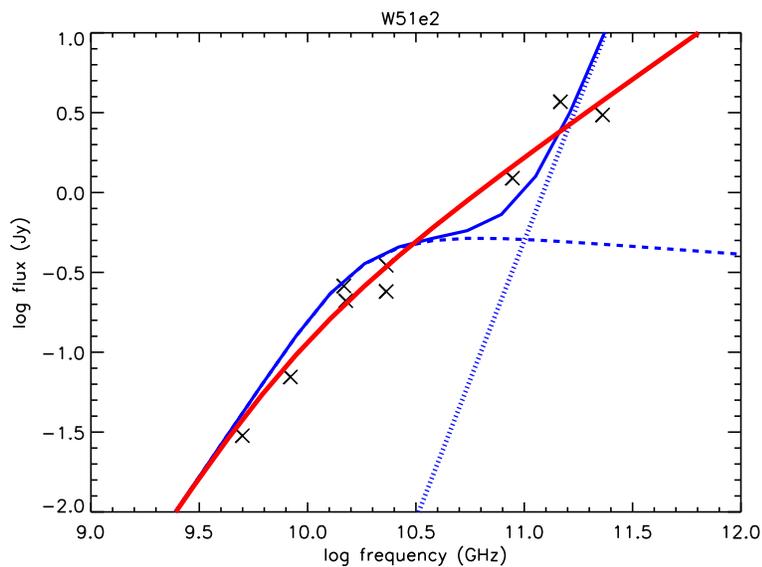}
\caption{
Measured flux densities of the W51e2 HII region (crosses). The blue dashed
lines show
a two-component model proposed by \citet{Rudolph1990} consisting of a 
uniform density HII region plus dust (dotted line).
The solid red line is the SED predicted for an HII region with
a density gradient ($n_e\propto r^{-1.5}$) and no significant dust emission.
See \S\ref{derivation} and table 7.
}
\label{fig:sedw51e2}
\end{figure}

\begin{figure}
\includegraphics[angle=90,width=4.truein]{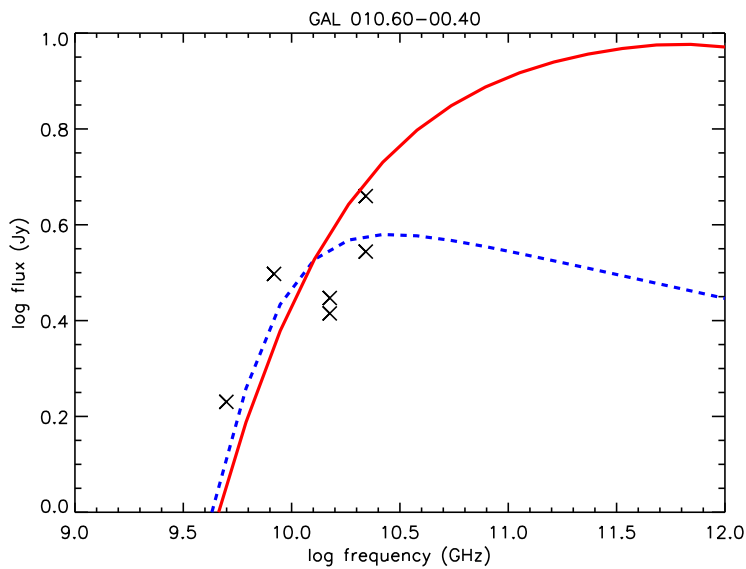}
\caption{Measured flux densities of the G10.6-0.4 HII region (crosses).
The blue dashed line is a model based on a uniform density HII region;
the red line is a model based on an HII region with a density gradient.}
\label{fig:sedG106}
\end{figure}

\begin{figure}
\includegraphics[angle=90,width=4.truein]{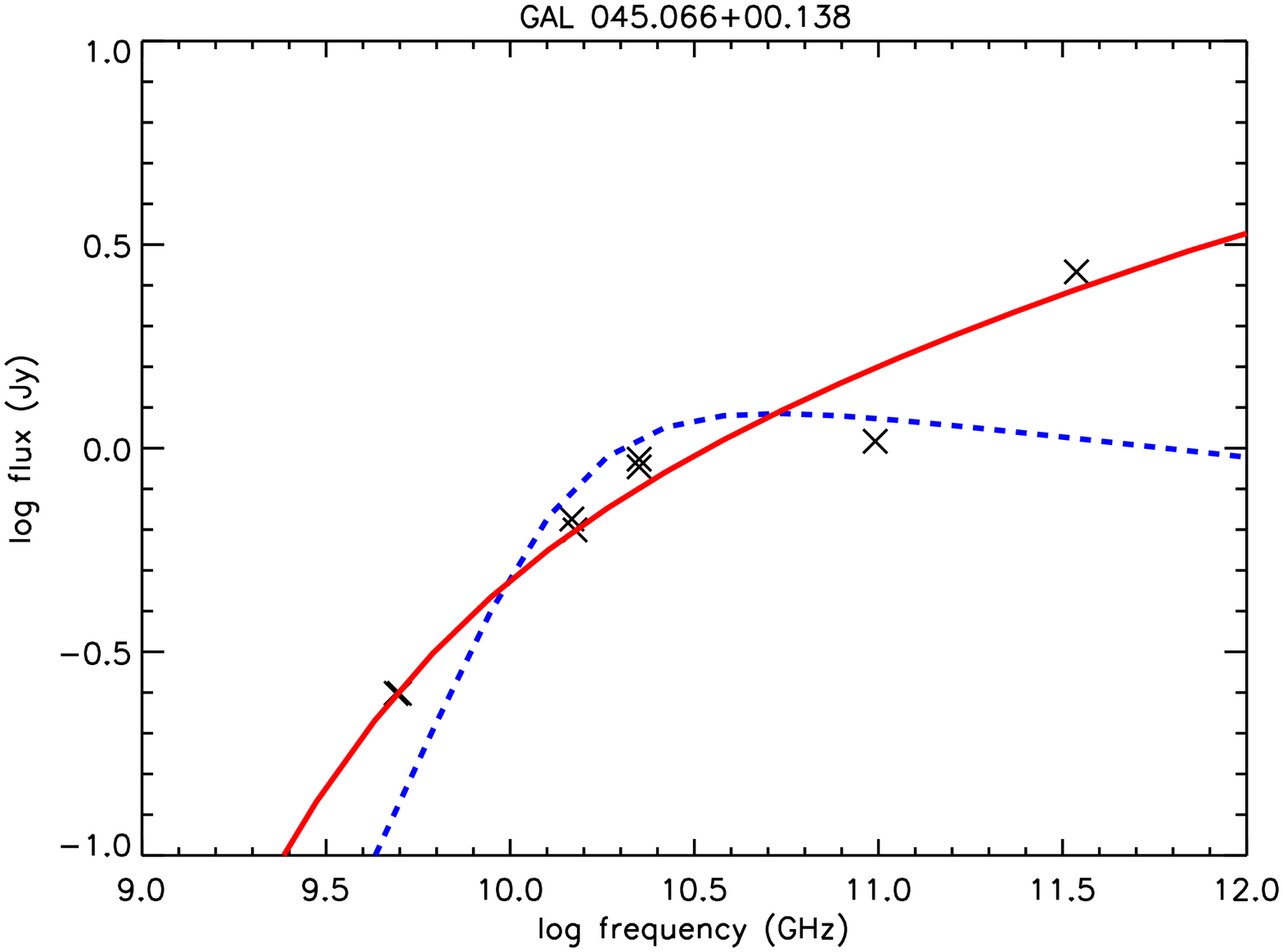}
\caption{ Continuum spectral energy distribution for G45.07+00.14.  
The crosses mark the observations; the blue dashed line
  is a model based on a uniform density HII region; the red line is a
  model based on an HII region with a density gradient.}
\label{fig:sedG45}
\end{figure}

\begin{figure}
\includegraphics[angle=90,width=4.truein]{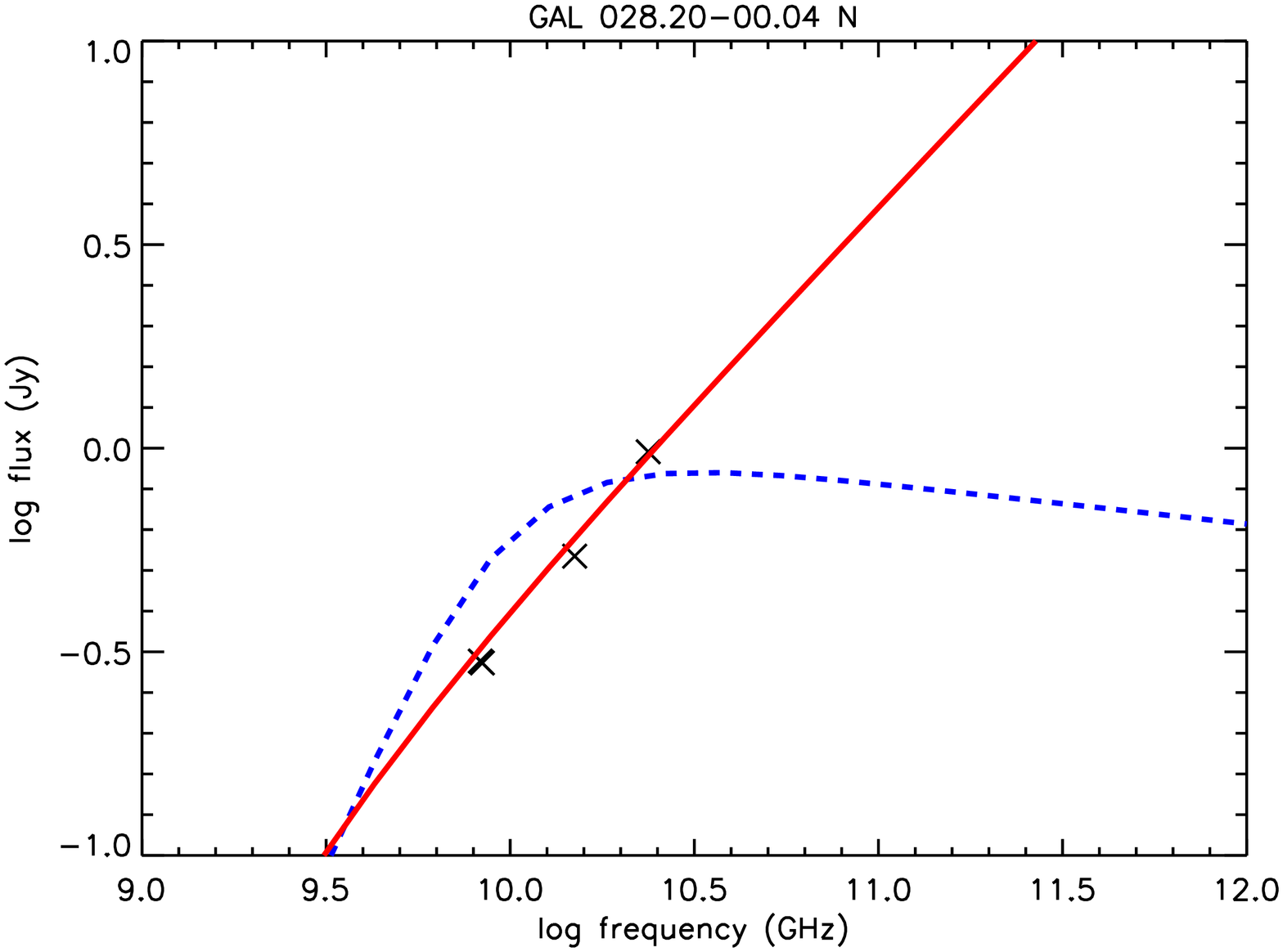}
\caption{ Continuum spectral energy distribution for G28.20-0.04.
  The crosses mark the observations; the blue dashed line is a model based
  on a uniform density HII region; the red line is a model based on
  an HII region with a density gradient.}
\label{fig:sedG28}
\end{figure}

\clearpage

\end{document}